%% file: main.tex
\documentclass[conference,10pt,a4paper]{IEEEtran}

\usepackage{cite}
\usepackage{comment}
\usepackage{amsmath}

\usepackage{xcolor,colortbl}
\definecolor{col1}{HTML}{3891A6}
\definecolor{col2}{HTML}{EF5B5B}
\definecolor{col3}{HTML}{3DDC97}

\usepackage{amsmath,amsthm,relsize}
\usepackage{amsfonts, amssymb, cuted}
\usepackage{cleveref}
\usepackage{mathtools}
\usepackage{algorithm}
\usepackage[noend]{algpseudocode}
\usepackage{cite}
\usepackage{soul}
\usepackage{tikz}
\usepackage{pgfplotstable}
\usepackage{pgfplots}
\usepackage{nicefrac}
\usepackage{enumitem}
\usepackage{support-caption}
\usepackage{subcaption}
\usepackage[font=footnotesize]{caption}
\usepackage{multirow}
\usepackage{relsize}
\usetikzlibrary{patterns}
\usepackage{acro}
\usepackage{pifont}
\usepackage{blkarray}

\usepackage{todonotes}

\usepackage{mathtools}
\usepackage{array}

\usepackage{diagbox}
\usepackage{smartdiagram}
\usesmartdiagramlibrary{additions}
\usepackage{bm}
\usepackage{multicol}

\usepackage{graphicx}  
\usepackage{tabularx}
\usepackage{colortbl}

\usepackage{color}
\usepackage{colortbl}
\usepackage{multirow}

\definecolor{intnull}{RGB}{213,229,255}
\definecolor{inteins}{RGB}{128,179,255}
\definecolor{intzwei}{RGB}{42,127,255}
\definecolor{intdrei}{RGB}{0,85,212}
\definecolor{intvier}{RGB}{0,51,128}
\definecolor{intfunf}{RGB}{0,34,85}

\input{commands}

\input{abbrev}

\begin{document}



\title{SIC-free Multicast Scheduling\\for Multi-antenna Coded Caching}
\vspace{-2mm}
\author{\IEEEauthorblockN{MohammadJavad Sojdeh, MohammadJavad Salehi, and Antti T\"olli} 
\IEEEauthorblockA{
    Centre for Wireless Communications (CWC), University of Oulu, 90570 Oulu, Finland \\
    \textrm{Emails:\{mohammadjavad.sojdeh, mohammadjavad.salehi, antti.tolli\}@oulu.fi}
    \vspace{-10pt}
    }
\thanks{
This work was supported by Infotech Oulu and by the Research Council of Finland under grants no. 343586 (CAMAIDE) and 369116 (6G Flagship).}
}
\maketitle
\input{abstract}
\input{Intro_ICASSP}

\input{SystemModel}

\input{Linear_scheme}
\input{SimulationResults}
\input{conclusions}
\clearpage
\bibliographystyle{IEEEtran}
\bibliography{conf_short,IEEEabrv,references,whitepaper}
\end{document}

%% file: commands.tex
\usepackage{arydshln}

\newcommand{\vbar}{\raisebox{.17ex}{\rule{.04em}{1.35ex}}}
\newcommand{\vbarind}{\raisebox{.01ex}{\rule{.04em}{1.1ex}}}
\newcommand{\R}{\ifmmode{\rm I}\hspace{-.2em}{\rm R} \else ${\rm I}\hspace{-.2em}{\rm R}$ \fi}
\newcommand{\T}{\ifmmode{\rm I}\hspace{-.2em}{\rm T} \else ${\rm I}\hspace{-.2em}{\rm T}$ \fi}
\newcommand{\N}{\ifmmode{\rm I}\hspace{-.2em}{\rm N} \else \mbox{${\rm I}\hspace{-.2em}{\rm N}$} \fi}
\newcommand{\B}{\ifmmode{\rm I}\hspace{-.2em}{\rm B} \else \mbox{${\rm I}\hspace{-.2em}{\rm B}$} \fi}
\newcommand{\Hil}{\ifmmode{\rm I}\hspace{-.2em}{\rm H} \else \mbox{${\rm I}\hspace{-.2em}{\rm H}$} \fi}
\newcommand{\C}{\ifmmode\hspace{.2em}\vbar\hspace{-.31em}{\rm C} \else \mbox{$\hspace{.2em}\vbar\hspace{-.31em}{\rm C}$} \fi}
\newcommand{\Cind}{\ifmmode\hspace{.2em}\vbarind\hspace{-.25em}{\rm C} \else \mbox{$\hspace{.2em}\vbarind\hspace{-.25em}{\rm C}$} \fi}
\newcommand{\Q}{\ifmmode\hspace{.2em}\vbar\hspace{-.31em}{\rm Q} \else \mbox{$\hspace{.2em}\vbar\hspace{-.31em}{\rm Q}$} \fi}
\newcommand{\Z}{\ifmmode{\rm Z}\hspace{-.28em}{\rm Z} \else ${\rm Z}\hspace{-.28em}{\rm Z}$ \fi}

\newtheorem{exmp}{Example}
\theoremstyle{definition}

\newtheorem{rem}{Remark}

\newcommand{\CE}[0]{{\mathcal{E}}}

\newcommand{\CG}[0]{{\mathcal{G}}}

\newcommand{\CK}[0]{{\mathcal{K}}}

\newcommand{\CP}[0]{{\mathcal{P}}}

\newcommand{\CS}[0]{{\mathcal{S}}}
\newcommand{\CT}[0]{{\mathcal{T}}}

\newcommand{\CW}[0]{{\mathcal{W}}}

\newcommand{\Ba}[0]{{\mathbf{a}}}

\newcommand{\Be}[0]{{\mathbf{e}}}

\newcommand{\Bh}[0]{{\mathbf{h}}}

\newcommand{\Bw}[0]{{\mathbf{w}}}
\newcommand{\Bx}[0]{{\mathbf{x}}}

\newcommand{\BA}[0]{{\mathbf{A}}}

\newcommand{\BN}[0]{{\mathbf{N}}}

\newcommand{\SfS}[0]{{\mathsf{S}}}

\usepackage{tikz}

\usepackage{tikz}
\usetikzlibrary{arrows,
                calc,
                decorations.pathreplacing,
                calligraphy,
                matrix,
                positioning
                }

%% file: abbrev.tex
\DeclareAcronym{ADMM}{
    short = ADMM,
    long = alternating direction method of multipliers,
    list = Alternating Direction Method of Multipliers,
    tag = abbrev
}

\DeclareAcronym{AoA}{
    short = AoA,
    long = angle-of-arrival,
    list = Angle-of-Arrival,
    tag = abbrev
}

\DeclareAcronym{SISO}{
    short = SISO,
    long = single-input single-output,
    list = single-input single-output,
    tag = abbrev
}

\DeclareAcronym{MRT}{
    short = MRT,
    long = maximum ratio transmitter,
    list = maximum ratio transmitter,
    tag = abbrev
}

\DeclareAcronym{PDA}{
    short = PDA,
    long = placement delivery array,
    list = placement delivery array,
    tag = abbrev
}

\DeclareAcronym{EE}{
    short = EE,
    long = energy efficiency,
    list = energy efficiency,
    tag = abbrev
}

\DeclareAcronym{MDS}{
    short = MDS,
    long = maximum distance separation,
    list = maximum distance separation,
    tag = abbrev
}

\DeclareAcronym{SIC}{
    short = SIC,
    long = successive-interference-cancellation,
    list = successive-interference-cancellation,
    tag = abbrev
}

\DeclareAcronym{MAC}{
    short = MAC,
    long = multiple-access-channel,
    list = multiple-access-channel,
    tag = abbrev
}

\DeclareAcronym{AoD}{
    short = AoD,
    long = angle-of-departure,
    list = Angle-of-Departure,
    tag = abbrev
}

\DeclareAcronym{BB}{
    short = BB,
    long = base band,
    list = Base Band,
    tag = abbrev
}

\DeclareAcronym{BC}{
    short = BC,
    long = broadcast channel,
    list = Broadcast Channel,
    tag = abbrev
}

\DeclareAcronym{BS}{
    short = BS,
    long = base station,
    list = Base Station,
    tag = abbrev
}

\DeclareAcronym{BR}{
    short = BR,
    long = best response,
    list = Best Response, 
    tag = abbrev
}

\DeclareAcronym{CB}{
    short = CB,
    long = coordinated beamforming,
    list = Coordinated Beamforming,
    tag = abbrev
}

\DeclareAcronym{CC}{
    short = CC,
    long = coded caching,
    list = Coded Caching,
    tag = abbrev
}

\DeclareAcronym{CE}{
    short = CE,
    long = channel estimation,
    list = Channel Estimation,
    tag = abbrev
}

\DeclareAcronym{CoMP}{
    short = CoMP,
    long = coordinated multi-point transmission,
    list = Coordinated Multi-Point Transmission,
    tag = abbrev
}

\DeclareAcronym{CRAN}{
    short = C-RAN,
    long = cloud radio access network,
    list = Cloud Radio Access Network,
    tag = abbrev
}

\DeclareAcronym{CSE}{
    short = CSE,
    long = channel specific estimation,
    list = Channel Specific Estimation,
    tag = abbrev
}

\DeclareAcronym{CSI}{
    short = CSI,
    long = channel state information,
    list = Channel State Information,
    tag = abbrev
}

\DeclareAcronym{CSIT}{
    short = CSIT,
    long = channel state information at the transmitter,
    list = Channel State Information at the Transmitter,
    tag = abbrev
}

\DeclareAcronym{CU}{
    short = CU,
    long = central unit,
    list = Central Unit,
    tag = abbrev
}

\DeclareAcronym{D2D}{
    short = D2D,
    long = device-to-device,
    list = Device-to-Device,
    tag = abbrev
}

\DeclareAcronym{DE-ADMM}{
    short = DE-ADMM,
    long = direct estimation with alternating direction method of multipliers,
    list = Direct Estimation with Alternating Direction Method of Multipliers,
    tag = abbrev
}

\DeclareAcronym{DE-BR}{
    short = DE-BR,
    long = direct estimation with best response,
    list = Direct Estimation with Best Response,
    tag = abbrev
}

\DeclareAcronym{DE-SG}{
    short = DE-SG,
    long = direct estimation with stochastic gradient,
    list = Direct Estimation with Stochastic Gradient,
    tag = abbrev
}

\DeclareAcronym{DFT}{
	short = DFT,
	long = discrete fourier transform,
	list = Discrete Fourier Transform,
	tag = abbrev
}

\DeclareAcronym{DoF}{
    short = DoF,
    long = degrees of freedom,
    list = Degrees of Freedom,
    tag = abbrev
}

\DeclareAcronym{DL}{
    short = DL,
    long = downlink,
    list = Downlink,
    tag = abbrev
}

\DeclareAcronym{GD}{
	short = GD, 
	long = gradient descent,
	list = Gradeitn Descent,
	tag = abbrev
}

\DeclareAcronym{IBC}{
    short = IBC,
    long = interfering broadcast channel,
    list = Interfering Broadcast Channel,
    tag = abbrev
}

\DeclareAcronym{i.i.d.}{
    short = i.i.d.,
    long = independent and identically distributed,
    list = Independent and Identically Distributed,
    tag = abbrev
}

\DeclareAcronym{JP}{
    short = JP,
    long = joint processing,
    list = Joint Processing,
    tag = abbrev
}

\DeclareAcronym{KKT}{
    short = KKT,
    long = Karush-Kuhn-Tucker,
    tag = abbrev
}

\DeclareAcronym{LOS}{
	short = LOS,
	long = line-of-sight,
	list = Line-of-Sight,
	tag = abbrev
}

\DeclareAcronym{LS}{
    short = LS,
    long = least squares,
    list = Least Squares,
    tag = abbrev
}

\DeclareAcronym{LTE}{
    short = LTE,
    long = Long Term Evolution,
    tag = abbrev
}

\DeclareAcronym{LTE-A}{
    short = LTE-A,
    long = Long Term Evolution Advanced,
    tag = abbrev
}

\DeclareAcronym{MIMO}{
    short = MIMO,
    long = multiple-input multiple-output,
    list = Multiple-Input Multiple-Output,
    tag = abbrev
}

\DeclareAcronym{MISO}{
    short = MISO,
    long = multiple-input single-output,
    list = Multiple-Input Single-Output,
    tag = abbrev
}

\DeclareAcronym{MSE}{
    short = MSE,
    long = mean-squared error,
    list = Mean-Squared Error,
    tag = abbrev
}

\DeclareAcronym{MMSE}{
    short = MMSE,
    long = minimum mean-squared error,
    list = Minimum Mean-Squared Error,
    tag = abbrev
}

\DeclareAcronym{mmWave}{
	short = mmWave,
	long = millimeter wave,
	list = Millimeter Wave,
	tag = abbrev
}

\DeclareAcronym{MU-MIMO}{
    short = MU-MIMO,
    long = multi-user \ac{MIMO},
    list = Multi-User \ac{MIMO},
    tag = abbrev
}

\DeclareAcronym{OTA}{
    short = OTA,
    long = over-the-air,
    list = Over-the-Air,
    tag = abbrev
}

\DeclareAcronym{PSD}{
    short = PSD,
    long = positive semidefinite,
    list = Positive Semidefinite,
    tag = abbrev
}

\DeclareAcronym{QoS}{
	short = QoS,
	long = quality of service,
	list = Quality of Service,
	tag = abbrev
}

\DeclareAcronym{RCP}{
	short = RCP,
	long = remote central processor,
	list = Remote Central Processor,
	tag = abbrev
}

\DeclareAcronym{RRH}{
    short = RRH,
    long = remote radio head,
    list = Remote Radio Head,
    tag = abbrev
}

\DeclareAcronym{RSSI}{
    short = RSSI,
    long = received signal strength indicator,
    list = Received Signal Strength Indicator,
    tag = abbrev
}

\DeclareAcronym{RX}{
	short = RX,
	long = receiver,
	list = Receiver,
	tag = abbrev
}

\DeclareAcronym{SCA}{
    short = SCA,
    long = successive convex approximation,
    list = Successive Convex Approximation,
    tag = abbrev
}

\DeclareAcronym{SG}{
    short = SG,
    long = stochastic gradient,
    list = Stochastic Gradient,
    tag = abbrev
}

\DeclareAcronym{SNR}{
    short = SNR,
    long = signal-to-noise ratio,
    list = Signal-to-Noise Ratio,
    tag = abbrev
}

\DeclareAcronym{SINR}{
    short = SINR,
    long = signal-to-interference-plus-noise ratio,
    list = Signal-to-Interference-plus-Noise Ratio,
    tag = abbrev
}

\DeclareAcronym{SOCP}{
	short = SOCP, 
	long = second order cone program,
	list = Second Order Cone Program,
	tag = abbrev
}

\DeclareAcronym{SSE}{
    short = SSE,
    long = stream specific estimation,
    list = Stream Specific Estimation,
    tag = abbrev
}

\DeclareAcronym{SVD}{
	short = SVD,
	long = singular value decomposition,
	list = Singular Value Decomposition,
	tag = abbrev
}

\DeclareAcronym{TDD}{
	short = TDD,
	long = time division duplex,
	list = Time Division Duplex,
	tag = abbrev
}

\DeclareAcronym{TX}{
	short = TX,
	long = transmitter,
	list = Transmitter,
	tag = abbrev
}

\DeclareAcronym{UE}{
    short = UE,
    long = user equipment,
    list = User Equipment,
    tag = abbrev
}

\DeclareAcronym{UL}{
    short = UL,
    long = uplink,
    list = Uplink,
    tag = abbrev
}

\DeclareAcronym{ULA}{
	short = ULA,
	long = uniform linear array,
	list = Uniform Linear Array,
	tag = abbrev
}

\DeclareAcronym{UPA}{
    short = UPA,
    long = uniform planar array,
    list = Uniform Planar Array,
    tag = abbrev
}

\DeclareAcronym{WMMSE}{
    short = WMMSE,
    long = weighted minimum mean-squared error,
    list = Weighted Minimum Mean-Squared Error,
    tag = abbrev
}

\DeclareAcronym{WMSEMin}{
    short = WMSEMin,
    long = weighted sum \ac{MSE} minimization,
    list = Weighted sum \ac{MSE} Minimization,
    tag = abbrev
}

\DeclareAcronym{WBAN}{
	short = WBAN,
	long = wireless body area network,
	list = Wireless Body Area Network,
	tag = abbrev
}

\DeclareAcronym{WSRMax}{
    short = WSRMax,
    long = weighted sum rate maximization,
    list = Weighted Sum Rate Maximization,
    tag = abbrev
}

%% file: abstract.tex

\begin{abstract}
Multi-antenna coded caching (CC) with multicast beamforming typically relies on a complex successive interference cancellation (SIC) structure to decode a superposition of multiple streams received by each user. Signal-level CC schemes require the regeneration and cancellation of interfering signals at the physical layer of each receiver, which complicates practical implementations.
To address this, we propose a bit-level multicast scheduling scheme enabling linear, SIC-free decoding of parallel streams by repeatedly transmitting data terms with linearly independent coefficients. Two reference strategies and a novel \emph{sparse} strategy are considered for constructing the coefficient matrix. The reference cases include the random strategy, which lacks control over matrix construction, and the equal-distant strategy, which balances users' interference and data terms equally.
In contrast, the sparse strategy minimizes the number of multicast streams transmitted in parallel during each interval. This approach simplifies both the decoding process and the beamforming design by decoupling the desired data terms for each user and reducing the number of SINR constraints, respectively.
To further enhance the symmetric rate, a successive projection algorithm is applied to exploit channel properties and optimize user ordering. With the coefficient matrix and optimized user ordering in place, multicast beamformers are devised to aggregate desired data from relevant multicast streams. Numerical simulations validate the effectiveness of the sparse strategy and user scheduling, demonstrating significant gains in symmetric rate.
\end{abstract}


%% file: Intro_ICASSP.tex
\section{Introduction}
\vspace{-1mm}
\label{section:intro}
Emerging new data-intensive services such as autonomous driving and mobile immersive viewing are putting wireless communication networks under mounting pressure to achieve higher data rates and lower latency~\cite{rajatheva2020white}. To achieve such requirements, novel coded caching (CC) schemes stand out for their intriguing potential to offer a performance boost proportional to the cumulative cache size of all network users~\cite{salehi2022enhancing}. 
Although CC was initially developed for single-input single-output (SISO) setups~\cite{maddah2014fundamental}, subsequent research revealed its applicability to multiple-input single-output (MISO) systems as well. This demonstrated that spatial multiplexing and CC gains are additive by serving multiple groups of users with multiple multicast messages and suppressing the intra-group interference by beamforming~\cite{shariatpanahi2018physical,salehi2020lowcomplexity,lampiris2021resolving}. Accordingly, in a MISO-CC setting with $L$ Tx antennas, $t + L$ users can be served in parallel, where the CC gain $t$ is proportional to the cumulative cache size of all users. However, this scheme necessitates each user to decode $\delta=\binom{t+L-1}{t}$ parallel streams in each transmission, resulting in a $\delta$-dimentional Gaussian multiple access channel (MAC). One well-known solution to decode parallel streams is to use the non-linear  successive interference cancellation (SIC) structure. However, SIC is undesired in practice as it requires complex receivers with just a marginal performance gain compared to simpler linear solutions~\cite{tolli2017multi,salehi2022multi}. Moreover,
as shown in \cite{mahmoodi2021low}, avoiding SIC can also significantly simplify the optimized beamformer design.

Eliminating SIC in the context of multi-antenna CC was first investigated in~\cite{tolli2017multi} by avoiding overlapping multicast groups, and thus removing MAC entirely. However, this solution is applicable only when $\frac{t+L}{t+1}$ is an integer. An alternative approach, explored in~\cite{lampiris2018adding}, involves combining codewords in the signal domain rather than adding data bits and utilizing a unicast (dedicated) beamformer for each data term. This solution not only allows us to avoid SIC but also reduces subpacketization through Placement Delivery Array (PDA) structure \cite{Yang-PDA-for-CC,Huang-PDA,Namboodiri-extended-PDA,Yan-PDA-for-centralized-CC}, paving the way for applying CC techniques in dynamic \cite{abolpour2023cache,Abolpour-twc,Abolpour_letter} and MIMO setups \cite{naseritehrani2023low,naseritehrani2024multicast}.
Signal-level CC schemes, however, suffer from inferior performance in the finite-SNR regimes due to the lack of multicast beamforming gain of the bit-level approach \cite{salehi2022multi,salehi2022enhancing} and require access to the cache memory at the physical layer~\cite{salehi2022enhancing}.


In this paper, a novel linear transmission solution is proposed to eliminate SIC while keeping the underlying bit-domain multicasting gain. Specifically, data terms are transmitted repeatedly, and linearly independent coefficient are used to guarantee decodability. In addition to reference strategies of a basic `Random' design and a fair `Equal-distance' strategy, we propose a maximally `Sparse' coefficient matrix construction
to minimize the number of data streams sent in each transmission. Such a sparse design improves performance as it allows more effective collection of the signal power and better interference cancellation.
Additionally, it simplifies the beamformer design by reducing the MAC size at each user and minimizing the number of interference terms to be suppressed~\cite {tolli2017multi}.
However, the underlying greedy user selection of the Sparse strategy imposes some \emph{penalty} on a subset of users in the general case as they have to decode part(s) of their requested data after multiple transmissions. Thus, an instantaneous channel-dependent successive projection algorithm is also applied to further enhance the performance. 
Given the coefficient matrix and optimized user ordering, multicast beamformers are designed to maximize the symmetric rate. Numerical simulations are presented to demonstrate the effectiveness of the proposed coefficient matrix construction and the user scheduling mechanism.

\textbf{Notation:}
Vectors and matrices are shown by boldface lower- and upper-case letters, respectively. $[K] \coloneq \{{1},...,{K}\}$, and $\{\mathbf{x}_{i}\}_{i \in \mathcal{I}}$ denotes the horizontal concatenation of all $\Bx_i$.

%% file: SystemModel.tex
\section{System Model}
\label{section:sys_model}
Downlink transmission from an $L$-antenna base station~(BS) to $K$ cache-enabled single-antenna users is considered. The BS has access to a library of $N$ files $\mathcal{W} = \{ W_1,...,W_N \}$ each of size $F$ bits. Each user $k \in [K]$ has a cache memory of size $MF$ bits.
The CC gain is defined as $t \triangleq \frac{KM}{N} \in \mathbb{N}$ and represents how many copies of the file library could be stored in the cache memories of all users. The system operation consists of two phases: placement and delivery. In the placement phase, users’ cache memories are filled with data. 
Similar to~\cite{maddah2014fundamental,shariatpanahi2018physical}, each file ${W}_{n}$ is split into $ \binom{K}{t}$ non-overlapping equal-sized subfiles $W_{n,\CP}$, where $\CP \subseteq [K],|\CP|= t$, and user $k$ caches all subfiles ${W}_{n,\mathcal{P}}$ for which $k \in \mathcal{P}$.

At the beginning of the delivery phase, each user $k \in [K]$ reveals its requested file $W_{d_k} \in \CW$.
Following the baseline MISO-CC scheme in~\cite{shariatpanahi2018physical}, 
the BS constructs and transmits (e.g., in consecutive time slots) a set of transmission vectors ${\mathbf{x}(i)} \in \mathbb{C}^L$, $i \in [\binom{K}{t+L}]$, where ${\mathbf{x}(i)}$ delivers parts of the requested data to every user in a subset $\mathcal{K}(i)$ of users with $|\mathcal{K}(i)| = t+L$. 
The transmission vector $\Bx(i)$ is built as
     $\Bx(i) = \sum_{\mathcal{T} \subseteq \CS^{\CK(i)} } \mathbf{w}_{\mathcal{T}} {X}_{\mathcal{T}}$,
where $\mathcal{S}^{\mathcal{K}(i)}=\{\mathcal{T} \subseteq \mathcal{K}(i),|\mathcal{T}|= t+1\}$ is the set of all multicast groups in  time slot $i$, ${X}_{\mathcal{T}}$ is the multicast message for the multicast group $\mathcal{T}$ chosen from a unit power complex Gaussian codebook, and
$\mathbf{w}_{\mathcal{T}} \in \mathbb{C}^{L}$ is the transmit beamformer for ${X}_{\mathcal{T}}$.
Consecutively, user $k \in \CK(i)$ receives
\begin{equation*}
    \begin{array}{ll}
         y_{k}(i) \!=\! \mathbf{h}_{k}^{H} \! \sum_{\mathcal{T} \in \mathcal{S}_{k}^{\mathcal{K}(i)}} \mathbf{w}_{\mathcal{T}} {X}_{\mathcal{T}} \!+ \! \mathbf{h}_{k}^{H}\! \sum_{\Bar{\mathcal{T}} \in \Bar{\mathcal{S}}_{k}^{\mathcal{K}(i)}}  \mathbf{w}_{\Bar{\mathcal{T}}} {X}_{\Bar{\mathcal{T}}}\! + \! z_{k}(i), 
    \end{array}
\end{equation*}
where $\Bh_k \in \mathbb{C}^{L}$ is the channel vector between the BS and user~$k$ (assumed to be perfectly known at the transmitter), ${z}_{k}(i) \sim \mathcal{CN}(0,\,N_0)\,$ is the additive white Gaussian noise at user $k$, and $\mathcal{S}_{k}^{\mathcal{K}(i)}=\{\mathcal{T} \in \mathcal{S}^{\mathcal{K}(i)}\mid k \in \mathcal{T}\}$ and $\Bar{\mathcal{S}}_{k}^{\mathcal{K}(i)}=\mathcal{S}^{\mathcal{K}(i)} \backslash \mathcal{S}_{k}^{\mathcal{K}(i)}$ are the sets of intended and interference multicast indices for user $k$, respectively. 
Since $|\mathcal{S}_{k}^{\mathcal{K}(i)}|= \delta$, the CC scheme in~\cite{shariatpanahi2018physical} requires each user to decode its intended data from a MAC of size $\delta$.

Let $T_i$ denote the duration (seconds) of time slot $i$ required so that all users decode their intended data from $\mathbf{x}(i)$. If $\theta$ is the length of each codeword (in bits), and if we define $R_i$ to be the max-min multicast rate (bits per second) at which the BS transmits a common message to all users in time interval $i$, then $T_i = \frac{\theta}{R_i}$, and the per-user symmetric rate is defined as $R_{sym} = \nicefrac{F}{\sum_{i}T_i} = \nicefrac{F}{\sum_{i}\frac{\theta}{R_i}}$.

%% file: Linear_scheme.tex
\section{SIC-free Multicast Scheduling for MISO-CC}
\label{section:Linear_solution}
The core idea behind our linear strategy is to split each time interval $i \in [\binom{K}{t+L}]$ into $\delta=\binom{t+L-1}{t}$ \emph{sub-intervals}, and in each sub-interval $d \in [\delta]$, transmit a linear combination of all the multicast terms in $\CS^{\CK(i)}$ (and their respective beamformers) as $\Bx(i,d) = \sum_{\CT \in \CS^{\CK(i)}} A_{\CT}(i,d) X_{\CT} \Bw_{\CT}$, where $A_{\CT}(i,d) \in \mathbb{C}$ denotes the coefficient assigned to the multicast term $X_{\CT}$ in sub-interval $d$ of interval $i$. For each interval $i$ and each $\CT \in \CS^{\CK(i)}$, let us define $\Ba_{\CT}(i)$ as a $\delta \times 1$ vector containing all coefficient $A_{\CT}(i,d)$. Let us also define $\BA(i) \triangleq \{\Ba_{\CT}(i)\}_{\CT \in \CS^{\CK}}$ $\in \mathbb{C}^{\delta \times |\CS^{\CK}|}$ as the matrix formed by the horizontal concatenation of all $\Ba_{\CT}(i)$ (the ordering is arbitrary). For notational simplicity, let us focus on a specific interval and drop the $i$ index (the same procedure is repeated in every interval).

After the transmission of $\Bx(d)$, user $k$ receives $y_k(d) = \Bh_k^H \Bx(d) + z_k(d)$, where $z_k(d)$ denotes the noise term at user $k$ in sub-interval $d$. After receiving $y_k(d)$ for all $d \in [\delta]$, user~$k$ has to solve the following system of linear equations to decode its requested data (i.e., $X_{\CT}$ for all $\CT \in \CS_{k}^{\CK}$): 
\begin{align}
    \mathbf{y}_{k} &=  \begin{bmatrix}
     y_k(1)   \\
     \vdots \\
     y_k(\delta) 
  \end{bmatrix} = \mathbf{A}_k \left[
\begin{array}{ccc}
   \mathbf{h}_{k}^{H} \mathbf{w}_{\CT_k^1} & \cdots &  0 \\
   \vdots & \ddots & \vdots \\
   0 & \cdots & \mathbf{h}_{k}^{H} \mathbf{w}_{\CT_k^{\delta}}
\end{array}
\right]\begin{bmatrix}
     {X}_{\CT_k^1}   \\
     \vdots \\
     {X}_{\CT_k^{\delta}} 
  \end{bmatrix}\nonumber\\
  &+ \mathbf{B}_k\left[
\begin{array}{ccc}
   \mathbf{h}_{k}^{H} \mathbf{w}_{\bar{\CT}_{k}^1} & \cdots &  0 \\
   \vdots & \ddots & \vdots \\
   0 & \cdots & \mathbf{h}_{k}^{H} \mathbf{w}_{\bar{\CT}_{k}^{\bar{\delta}}}
\end{array}
\right]\begin{bmatrix}
     X_{\bar{\CT}_{k}^{1}}   \\
     \vdots \\
     {X}_{\bar{\CT}_{k}^{\bar{\delta}}}
  \end{bmatrix}+ \begin{bmatrix}
      z_k(1)  \\
      \vdots \\
      z_k(\delta)
  \end{bmatrix},
\label{system_of_equations}
\end{align}
where $\bar{\delta} = |\bar{\CS}_k^{\CK}| = \binom{t+L-1}{t+1}$ is the number of interference terms for user~$k$, $\CT_k^n$ and $\bar{\CT}_{k}^{m}$ show the $n$-th and $m$-th element of $\CS_k^{\CK}$ and $\bar{\CS}_k^{\CK}$, respectively (any ordering can be used), $\BA_k \triangleq \{\mathbf{a}_{\mathcal{T}}\}_{\mathcal{T} \in {\mathcal{S}}_{k}^{\mathcal{K}}}$ $\in \mathbb{C}^{\delta \times \delta}$, and $\mathbf{B}_k \triangleq \{\mathbf{a}_{\mathcal{T}}\}_{\mathcal{T} \in {\bar{\mathcal{S}}}_{k}^{\mathcal{K}}}$ $\in \mathbb{C}^{\delta \times \bar{\delta}}$.

Assume user~$k$ can decode all its required terms with the equal rate\footnote{Symmetric rate is imposed to minimize the time needed to receive all parallel messages.} of ${R}_{k}^{LIN} = \min_{n \in [\delta]} R_k^n$, where $R_k^n$ denotes the achievable rate for decoding $X_{\CT_k^n}$.
As the size of each codeword is $\theta$ and all the codewords are decoded in parallel, the time required for user~$k$ to decode its requested data is $T_k = \nicefrac{\theta}{\delta{R}_{k}^{LIN}}$ (seconds), and the total delivery time in all sub-intervals will be $T = \nicefrac{\theta}{\min_{k \in \CK} {R}_{k}^{LIN}}$. Now, we can write the symmetric rate maximization problem as
\begin{align}
&\max_{\gamma_{k}^{n},\mathbf{w}_{\CT},\mathbf{A}}  \min_{k \in \CK,n \in [\delta]}  R_k^n \nonumber\\
 &\textrm{s.t.} \hspace{4mm} {R}_{k}^{n} \le \log(1+\gamma_{k}^{n}), \nonumber\\
  & {\gamma}_{k}^{n} \le \frac{{\lvert \det(\mathbf{A}_{k}) \mathbf{h}_{k}^{H} \mathbf{w}_{\CT_k^n}
\rvert}^2}
{\sum\limits_{m \in [\bar{\delta}]} {\lvert \det(\mathbf{B}_{n,\bar{\CT}_k^{m}}) \mathbf{h}_{k}^{H} \mathbf{w}_{\bar{\CT}_k^{m}} \rvert}^2 + N_0\sum\limits_{j\in [\delta]} {\lvert \det(\mathbf{N}_{n,j}) \rvert}^2}, \nonumber\\
&\qquad\qquad\qquad\qquad\qquad\qquad\qquad\qquad\forall k \in \CK, \hspace{2mm} n \in [\delta],\nonumber\\
  & {\sum_{d \in [\delta]} \sum_{\CT \in \mathcal{S}^{\mathcal{K}}} {\lVert A_{\CT}(d) \mathbf{w}_{\mathcal{T}} \rVert}^{2}} \le \delta \times P_T,
  \label{Non_convex_optimization}
\end{align}
\textcolor{black}{where $\mathbf{B}_{n,\bar{\CT}_k^m} \triangleq \{\Ba_{\CT}\}_{\CT \in (\CS_k^{\CK} \backslash \CT_k^n)\cup \bar{\CT}_k^m}$ is the stream-specific interference matrix built by replacing the $n$-th column of $\BA_k$ with the $m$-th column of $\mathbf{B}_k$, and $\BN_{n,j} \in \mathbb{C}^{(\delta-1) \times (\delta-1)}$ is the noise amplification matrix, obtained by removing the $j$-th row of the matrix $\{\Ba_{\CT}\}_{\CT \in \CS_k^{\CK} \backslash \CT_k^n}$ (in other words, the term $N_0\sum_{j\in [\delta]} {\lvert \det(\mathbf{N}_{n,j}) \rvert}^2$ characterizes the noise aggregation over all sub-intervals)}. In the optimization problem \eqref{Non_convex_optimization}, variables $\mathbf{w}_{\CT}$ and $\mathbf{A}$ are coupled, resulting in a complex non-convex problem. To cope with this, one approach could be decoupling the variables and optimizing them in an alternative fashion. For fixed $\BA$, the problem remains non-convex w.r.t. $\mathbf{w}_{\CT}$ but can be tightly bounded and approximated using successive convex approximation (SCA). However, finding optimum $\BA$ for given $\Bw_{\CT}$ is still challenging as the summation over the $\det(\cdot)$ function is neither convex nor concave.
To address this issue, in the next section, we propose an offline 
selection method for obtaining $\mathbf{A}$.

Now, let us review SCA for finding $\Bw_{\CT}$ (for a given $\BA$). First, we rewrite~\eqref{Non_convex_optimization} in an epigraph form as
\begin{align}
& \max\limits_{r, {\gamma}_{k}^{n},\mathbf{w}_{\mathcal{T}}} r \nonumber\\
\textrm{s.t.} \quad &r \le \log(1+\gamma_{k}^{n}), \hspace{1mm} \forall k \in \mathcal{K}, \hspace{1mm} n \in [\delta],\nonumber\\
&\textrm{The rest of the constraints as in \eqref{Non_convex_optimization}},
\end{align}
which is non-convex due to the non-convex SINR constraints in~\eqref{Non_convex_optimization}. Following the same approach as in~\cite{tolli2017multi}, we rearrange the SINR condition and use first-order Taylor approximation to write the approximated convex function in the epigraph form:
\begin{equation}
\begin{aligned}
\max\limits_{r, {\gamma}_{k}^{n},\mathbf{w}_{\mathcal{T}}} &r \nonumber\\
\textrm{s.t.} \quad &r \le \log(1+\gamma_{k}^{n}),\\
 & \mathcal{L}(\mathbf{w}_{\CT_k^n}, \mathbf{w}_{\bar{\CT}_k^{m}},\mathbf{h}_{k},\gamma_{k}^{n}) \geq  \\
 &\;\;{\sum_{m \in [\bar{\delta}]} {\lvert \det(\mathbf{B}_{n,\bar{\CT}_k^{m}}) \mathbf{h}_{k}^{H} \mathbf{w}_{\bar{\CT}_k^{m}} \rvert}^2 + N_0\sum_{j\in [\delta]} {\lvert \det(\mathbf{N}_{n,j})\rvert}^2}\!\!\!,\nonumber\\
 & \qquad \qquad \qquad \qquad \qquad \qquad \qquad \quad \forall k \in \mathcal{K}, \hspace{1mm} n \in [\delta],\nonumber\\
& {\sum_{d \in [\delta]} \sum_{\CT \in \mathcal{S}^{\mathcal{K}}} {\lVert A_{\CT}(d) \mathbf{w}_{\mathcal{T}} \rVert}^{2}} \le \delta \times P_T,
\label{convexified-problem}
\end{aligned}
\end{equation}
{where operator $\mathcal{L}(.)$ denotes the first-order Taylor approximation of its input arguments (w.r.t the optimal point at previous iteration) as defined in \cite{tolli2017multi}.}

\begin{exmp}

Consider a content delivery scenario where a BS with $L=2$ antennas delivers requested files to $K=3$ users from a library $\mathcal{W}=\{A,B,C\}$ of $N=3$ files, each of size $F$ bits. The CC gain is $t=1$, so in the placement phase, each file is split into $3$ sub-files (e.g., $A \rightarrow \{A_1,A_2,A_3\}$), and each user~$k$ stores the $k$-th sub-file of every file (e.g., user~1 stores $A_1$, $B_1$, and $C_1$).
In the content delivery phase, suppose users 1, 2, and 3 request $A$, $B$, and $C$, respectively. As $t=1$, the set of all multicast groups is $\mathcal{S}^{\mathcal{K}}=\{\{1,2\},\{1,3\},\{2,3\}\}$.
As $\delta = 2$, we need 2 transmission vectors as
    $\mathbf{x}(1) = {A}_{12}(1) \mathbf{w}_{12} {X}_{12} + {A}_{13}(1) \mathbf{w}_{13} {X}_{13} + {A}_{23}(1) \mathbf{w}_{23} {X}_{23}$ and 
    $\mathbf{x}(2) = {A}_{12}(2) \mathbf{w}_{12} {X}_{12} + {A}_{13}(2) \mathbf{w}_{13} {X}_{13} + {A}_{23}(2) \mathbf{w}_{23} {X}_{23}$, 
where the brackets and element separators are removed from multicast sets for notation simplicity.
Accordingly, user $1$ needs to solve the linear system of equations
\begin{align}
    {y}_{1}(1) =  & \Big[\Big(\mathbf{h}_{1}^{H}\mathbf{w}_{12} \Big)\mathbf {A}_{12}(1) {X}_{12} +  \Big(\mathbf{h}_{1}^{H}\mathbf{w}_{13} \Big)\mathbf {A}_{13}(1) {X}_{13} \nonumber\\
    &+ \Big(\mathbf{h}_{1}^{H}\mathbf{w}_{23} \Big)\mathbf {A}_{23}(1) {X}_{23} +{z}_{1}(1)\Big],\nonumber
    \end{align}
    \begin{align}
    {y}_{1}(2) =    &\Big[\Big(\mathbf{h}_{1}^{H}\mathbf{w}_{12} \Big)\mathbf {A}_{12}(2) {X}_{12} +  \Big(\mathbf{h}_{1}^{H}\mathbf{w}_{13} \Big)\mathbf {A}_{13}(2) {X}_{13}\nonumber\\
    &+ \Big(\mathbf{h}_{1}^{H}\mathbf{w}_{23} \Big)\mathbf {A}_{23}(2){X}_{23}+{z}_{1}(2)\Big], \nonumber
\end{align}
to decode its requested terms $X_{12}$ and $X_{13}$. It can be followed that the SINR for decoding $X_{12}$ at user~1 is bounded by
\begin{equation}
    \gamma_{1}^{1} \le \frac{\lvert \mathbf ({A}_{12}(1){A}_{13}(2) - {A}_{12}(2){A}_{13}(1) ) \mathbf{h}_{1}^{H} \mathbf{w}_{12} \rvert^2}{\lvert \mathbf ({A}_{13}(1){A}_{23}(2) - {A}_{13}(2){A}_{23}(1) ) \mathbf{h}_{1}^{H} \mathbf{w}_{23} \rvert^2 + z_1^1},
    \label{example_sinr}
\end{equation}
where $z_1^1 \triangleq ({\lvert {A}_{13}(1) \rvert}^2+{\lvert {A}_{13}(2) \rvert}^2)N_0$. The coefficients in the numerator and denominator of~\eqref{example_sinr} are equivalent to
\begin{equation*}
    \det\left(\begin{bmatrix}
        A_{12}(1) &A_{13}(1) \\
        A_{12}(2) &A_{13}(2)
    \end{bmatrix}\right), \quad
    \det\left(\begin{bmatrix}
        A_{23}(1) &A_{13}(1) \\
        A_{23}(2) &A_{13}(2)
    \end{bmatrix}\right),
\end{equation*}
respectively, and the noise power $N_0$ is multiplied by $\det^2([A_{13}(1)]+[A_{13}(2)])$ in $z_1^1$. Similar expressions can be derived for $\gamma_1^2$ (i.e., the SINR term for decoding $X_{13}$ at user~1), and also for other SINR terms at users~2 and~3. 
\end{exmp}

\section{Designing the Coefficient Matrix}
The coefficient matrix $\BA$ (with dimensions $\delta \times |\CS^{\CK}|$) significantly impacts system performance as it alters both the numerator and the denominator of the SINR terms in~\eqref{Non_convex_optimization}.
\subsection{Reference Strategies}
\noindent\textbf{Random Generation}:
The first reference strategy is to randomly select each element of $\BA$ according to a given distribution. For example, one may use uniform distribution over a finite field $\mathbb{F}_q$ and choose the field size $q$ large enough to ensure each user-specific sub-matrix $\BA_k$ is invertible with high probability.  While this random generation approach is straightforward, it is generally expected to perform poorly because it does not exploit the specific structure of the problem.

\noindent\textbf{Equal-Distance Generation}: This approach constructs the multicast coefficient vectors $\Ba_{\CT}$ such that the inner product $\langle \mathbf{a}_{\mathcal{T}_1}, \mathbf{a}_{\mathcal{T}_2} \rangle$ is the same for all pairs of multicast groups $ \mathcal{T}_1, \mathcal{T}_2 \in \mathcal{S}^{\mathcal{K}}$. 
The intuition here is to make the signal and interference terms in~\eqref{Non_convex_optimization} to scale with the same factor, i.e., $\det(\BA_k)=\det(\mathbf{B}_{n,\bar{\CT}_k^{m}})= c$, for all $n \in [\delta]$, $m \in [\bar{\delta}]$, and for all users, where the constant $c$ depends on the size of the matrix.
The strength of this method is in its \emph{fairness}, as it guarantees the same effect on each user~$k \in \CK$. However, its drawback is that it forces all the signal and interference multipliers to be equal while it is possible to increase the signal multipliers and decrease the interference multipliers with a more intelligent, user-centric design. For example, one could constraint only the subsets of vectors intended for each user~$k$ by setting $\langle \mathbf{a}_{\mathcal{T}_1}, \mathbf{a}_{\mathcal{T}_2} \rangle$ to be the same for all pairs of multicast groups $ \mathcal{T}_1, \mathcal{T}_2 \in \mathcal{S}_{k}^{\mathcal{K}}$.

Generating equal-distance vectors closely follows the problem of \textit{Equiangular Tight Frames} (ETFs), where the goal is to find a set of unit-norm vectors such that the absolute value of the inner product of any pair of vectors is identical and minimal. Solving this problem, which is also known as Welch-bound-equality sequences in the literature, typically involves a challenging non-convex min-max optimization problem, and several methods exist for their construction under specific constraints~\cite{ETFS_sensing,ETFs_projection,ETFs_monotonic}.

\subsection{Sparse Generation}
The goal of the third (main) approach is to make each user-specific $\BA_k$ as close as possible to the identity matrix while ensuring the decodability constraint (i.e., $\det(\BA_k) \neq 0$, $\forall k \in \CK$). Clearly, this results in a sparse $\BA$ as many coefficients will be set to zero. The intuition behind this design is to increase the signal multiplier (i.e., $\det(\BA_k)$) in~\eqref{Non_convex_optimization}.
Additionally, unlike the equal-distance method, this user-centeric approach allows repeating multicast coefficient vectors, such as $\Ba_{\CT_1} = \Ba_{\CT_2}$, as long as $\CT_1 \cap \CT_2 = \varnothing$. Such repetitions are desired, as when $\Ba_{\CT_1} = \Ba_{\CT_2}$, the interference caused by $X_{\CT_2}$ over $X_{\CT_1}$ (and vice versa) is nulled out as the determinant of the respective interference matrices (i.e., $\mathbf{B}_{n,\bar{\CT}_k^m}$) become zero. As a result, SINR will be improved by increasing the signal multiplier and reducing the interference multiplier.

Moreover, the sparse approach simplifies both the decoding process and the beamformer design. 
Decoding is simplified because the data streams intended for each user are decoupled from each other (each zero entry in $\mathbf{A}$ means the multicast signal is \emph{not} transmitted in the respective sub-interval). 
Furthermore, the beamformer design is simplified as the sparse design effectively reduces the MAC size at each user, thus reducing the number of sum-rate SINR constraints in the optimized beamformer design problem (c.f~\cite{tolli2017multi}).

\begin{rem}
The $\beta$-scheme in~\cite{tolli2017multi} for removing the need for SIC processing ($\beta=1$) is equivalent to the proposed sparse matrix generation in a special case when $\frac{t+L}{t+1}$ is an integer. In fact, for integer $\frac{t+L}{t+1}$, it can be shown that the coefficient matrix $\BA$ can be built such that each user-specific sub-matrix $\BA_k$ is reduced to the identity matrix, and decoding each data stream becomes possible independently of other streams. On the other hand,
for the general case of non-integer $\frac{t+L}{t+1}$, one can show that it is \emph{not} possible to find a coefficient matrix such that each $\BA_k$ is an identity matrix. So, sparse matrix generation will impose some \emph{penalty} on a subset (or all) of users as they have to decode part(s) of their requested data after multiple sub-intervals. Indeed, one could distribute this decoding penalty over different users in various ways, and hence there is no unique way to generate a sparse $\BA$.
\end{rem}
In the following, a greedy approach is presented for generating a sparse coefficient matrix $\BA$.
Ideally, $\BA$ should be designed such that each user-specific coefficient matrix \(\mathbf{A}_{k}\) is an identity matrix of size $\delta \times \delta$. However, this is only feasible when $\frac{t+L}{t+1}$ is an integer~\cite{tolli2017multi}.
Instead, the proposed algorithm aims to assign a vector with the fewest `1' elements to each multicast group while meeting the linear decoding criteria. This ensures the final matrix $\BA$ is as sparse as possible.
The algorithm is executed in multiple iterations, where at each iteration, a new column of $\BA$ is assigned from the set of the standard basis vectors of the $\delta$-dimensional vector space and all their combinations. The pseudo-code of our proposed solution is provided in Algorithm~\ref{alg:cap}. Here, we provide a brief step-by-step explanation.


\begin{algorithm}[t]
\label{alg:main}
\small
\caption{Sparse Matrix Generation}
\label{alg:cap}
\begin{algorithmic}[1]
    \While{there is at least one active user}
        \State \label{algline:userselect}$k \gets $ the active user with the highest priority
        \State \label{algline:mcsets}$\hat{\CS}_k^{\CK} \gets \{\CT \in \SfS^{\CK}_k \mid \CT \textrm{ is active}\}$
        \ForAll{$\CT \in \hat{\CS}_k^{\CK}$} \label{algline:candset_begin}
            \State \label{algline:candset_end} $\CG_{\CT} \gets \CE \; \backslash \; \cup_{k \in \CT} \bar{\CE}_k$
            \State $\Ba_{\CT} \gets$ the assigned vector to $\CT$ 
            \ForAll{$k' \in \CT$, $k'$ is active}
                    \ForAll{$\Be \in \bar{\CE}_{k'}$}
                        \State Add $\Be + \Ba_{\CT}$ to $\bar{\CE}_{k'}$
                    \EndFor
                \State Add $\Ba_{\CT}$ to $\bar{\CE}_{k'}$
            \EndFor
            \State Mark $\CT$ as inactive
        \EndFor
        \State \label{algline:update_begin}Mark user~$k$ as inactive
        \EndWhile
\end{algorithmic}
\end{algorithm}

\begin{figure}[t]
\centering
\captionsetup{aboveskip=2pt, belowskip=0pt}
    \scalebox{0.8}{\input{Figs/Plot_4_javad}}
    \caption{Comparison with SIC and No-CC: $K=5, L=4, t=1$}
    \vspace{-5pt}
    \label{fig:SIC_5_user}
\end{figure}
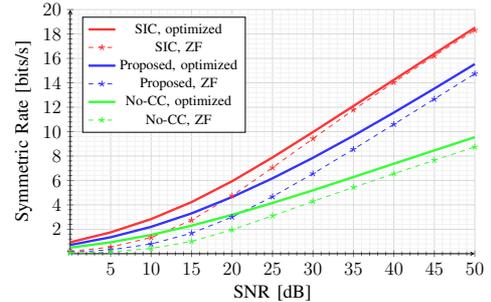

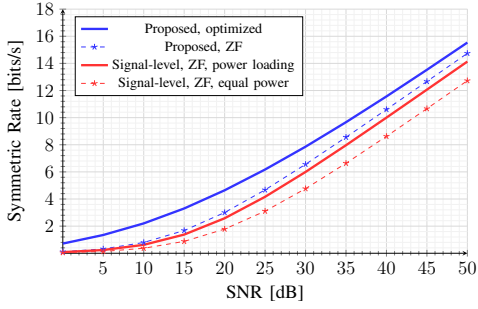
\begin{figure}[t]
\centering
\captionsetup{aboveskip=2pt, belowskip=0pt}
    \scalebox{0.8}{\input{Figs/Plot_3_javad}}
    \caption{Comparison with Signal-level CC: $K=5, L=4, t=1$}
    \label{fig:Signal_level}
\end{figure}

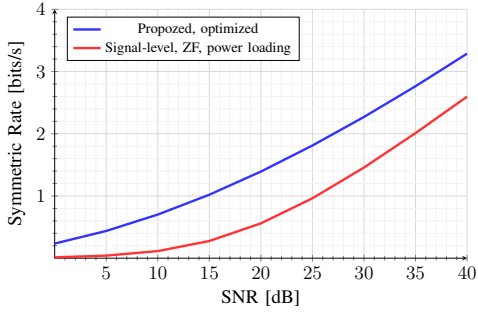
\begin{figure}[t]
\centering
\captionsetup{aboveskip=2pt, belowskip=0pt}
   \scalebox{0.8}{\input{Figs/Plot_7_javad}}
    \caption{Scalability of the proposed scheme: $K=20, L=6, t=1$.}
    \vspace{-6pt}
    \label{fig:20-users}
\end{figure}

\begin{figure}[t]
\centering
\captionsetup{aboveskip=2pt, belowskip=0pt}
    \scalebox{0.8}{\input{Figs/Plot_5_javad}}
    \caption{Coefficient matrix effect: $K=5, L=4, t=1$}
\label{fig:coefficient_matrix}
\end{figure}
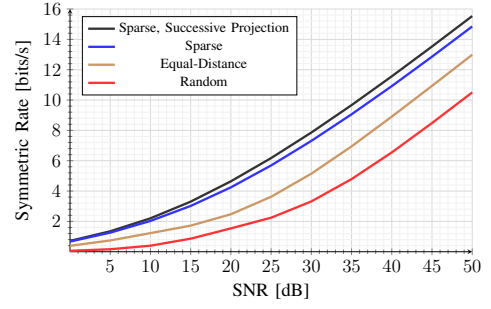

\noindent \textbf{User selection}: In each iteration, the algorithm selects a user~$k$ with the highest priority among the active users in $\CK$, and then, constructs $\hat{\CS}_k^{\CK}$ to include all active multicast groups containing user \(k\).\footnote{User~$k$ is active only if the algorithm has not yet selected it, and a multicast group $\CT$ is active only if its respective column in $\BA$ has not yet been assigned by the algorithm.} It then proceeds by selecting the first element of $\hat{\CS}_k^{\CK}$.

\noindent \textbf{Building candidate set:} 
In this step, the algorithm identifies a set of candidate vectors, $\CG_{\CT}$, that can be assigned to the multicast group $\CG_{\CT}$ without violating decodability conditions. To facilitate this, we define the standard basis vectors of the $\delta$-dimensional vector space as $\Be_l,{l \in [\delta]}$. From these basis vectors, we construct the ordered set $\CE$, which includes all possible summations of one or more $\Be_l$. The elements in $\CE$ are ordered by the number of their `1' elements. Finally, candidate set $\CG_{\CT}$ is obtained as $\CG_{\CT} = \CE \; \backslash \; \cup_{k \in \CT} \bar{\CE}_k$, where $\bar{\CE}_k$ explicitly includes all vectors that must \emph{not} be assigned to any multicast group that includes user~$k$.


\noindent \textbf{Vector assignment:} 
Since the elements of $\CG_{\CT}$ are prioritized, the first vector in $\CG_{\CT}$ is selected as the assigned vector for the multicast group.

\noindent \textbf{Parameter updates:}
Given the vector assignment, multicast group~$\CT$ is marked inactive, and its corresponding column in $\BA$ (i.e., $\Ba_{\CT}$) is filled with its assigned vector. Moreover, for each active user $k' \in \CT$, $\Ba_{\CT}$ and its summation with each element of $\bar{\mathcal{E}}_{k'}$ is added to $\bar{\mathcal{E}}_{k'}$. This is done to avoid the repetition of the columns (and so, the linear dependence) in $\BA_{k'}$. Finally, the algorithm is repeated until all the multicast groups of the users are inactive.

\begin{exmp}
Assume a MISO setup with $L=4$ antennas at the transmitter and the CC gain of $t=1$. For this setup, consider a time interval where the set of users $\CK = \{1,2,3,4,5\}$ is served by multicast transmissions of codewords of size $t+1 = 2$. Assume the users in $\CK$ are prioritized by their index, such that the user~1 has the highest priority.

As $\delta=4$, the set of all possible coefficient vectors is ordered as $\CE = \{\Be_1,\cdots,\Be_4,\Be_1+\Be_2,\cdots,\Be_3+\Be_4,\Be_1+\Be_2+\Be_3,\cdots,\Be_2+\Be_3+\Be_4,\Be_1+\Be_2+\Be_3+\Be_4\}$.

In the first iteration, Algorithm~\ref{alg:cap} starts with user~$1$, for which
the set of active multicast groups $\hat{\CS}_1^{\CK}=\{\{1,2\},\{1,3\},\{1,4\},\{1,5\}\}$.
Since $\bar{\CE}_1= \bar{\CE}_2=\varnothing$, $\mathcal{G}_{12} =\CE$. It can be easily verified that Algorithm~\ref{alg:cap} iterates over all multicast groups of user~$1$ to perform the vector assignment as: $\Ba_{12}=\Be_1$, $\Ba_{13}=\Be_2$, $\Ba_{14}=\Be_3$, $\Ba_{15}=\Be_4$. Then, $\bar{\CE}_{k}$ sets are updated as $\bar{\CE}_2=\{\Be_1\}$, $\bar{\CE}_3=\{\Be_2\}$, $\bar{\CE}_4=\{\Be_3\}$, $\bar{\CE}_5=\{\Be_4\}$.

In the next iteration, Algorithm~\ref{alg:cap} selects user~2, for which the set of active multicast groups is $\hat{\CS}_2^{\CK}=\{\{2,3\},\{2,4\},\{2,5\}\}$.
Given $\bar{\CE}_k$ sets from the previous iteration, the candidate set  $\mathcal{G}_{23}$ is calculated as $\mathcal{G}_{23} = \CE \backslash \{\Be_1,\Be_2\}$. Repeating the same procedure for all multicast groups of user~2 results in the following vector assignments: $\Ba_{23}=\Be_3$, $\Ba_{24}=\Be_4$, $\Ba_{25}=\Be_2$. Then, $\bar{\CE}_{k}$ sets are updated as $\bar{\CE}_3=\{\Be_2,\Be_3,\Be_2+\Be_3\}$, $\bar{\CE}_4=\{\Be_3,\Be_4,\Be_3+\Be_4\}$, $\bar{\CE}_5=\{\Be_4,\Be_2,\Be_4+\Be_2\}$.

For user~3, we have $\hat{\CS}_3^{\CK}=\{\{3,4\},\{3,5\}\}$. The candidate set $\mathcal{G}_{34}$ is determined as  $\mathcal{G}_{34} = \CE \backslash \{\Be_2,\Be_3,\Be_4,\Be_2+\Be_3,\Be_3+\Be_4\}$. The algorithm assigns the vectors as $\Ba_{34} = \Be_1$ and $\Ba_{35} = \Be_1+\Be_4$. Subsequently, the sets $\bar{\CE}_{k}$ are updated to $\bar{\CE}_4=\{\Be_1,\Be_3,\Be_4,\Be_1+\Be_3,\Be_1+\Be_4,\Be_3+\Be_4,\Be_1+\Be_3+\Be_4\}$, $\bar{\CE}_5=\{\Be_2,\Be_4,\Be_1+\Be_4,\Be_2+\Be_4,\Be_1+\Be_2+\Be_4\}$.

Following the same process for users~4 and~5, the sparse coefficients matrix $\BA$ is calculated as:
\[
\fontsize{8pt}{10pt}\selectfont
\begin{array}{rl}
\mathbf{A} = &
\begin{array}{l}
\end{array} 
\left[
\begin{array}{*{10}{@{\hskip 0.05cm}c@{\hskip 0.05cm}}}\mathbf{a}_{12} & \mathbf{a}_{13} & \mathbf{a}_{14} & \mathbf{a}_{15} & \mathbf{a}_{23} & \mathbf{a}_{24} & \mathbf{a}_{25} & \mathbf{a}_{34} & \mathbf{a}_{35} & \mathbf{a}_{45} \\
1 & 0 & 0 & 0 & 0 & 0 & 0 & 1 & 1 & 0 \\
0 & 1 & 0 & 0 & 0 & 0 & 1 & 0 & 0 & 1 \\
0 & 0 & 1 & 0 & 1 & 0 & 0 & 0 & 0 & 1 \\
0 & 0 & 0 & 1 & 0 & 1 & 0 & 0 & 1 & 0 \\
\end{array}
\right] 
\end{array}
\]
\end{exmp}
Each user-specific coefficients matrix $\BA_k$ is built by horizontal concatenation of all multicast coefficient vectors $\Ba_{\CT}$ that contain user~$k$. For instance, $\BA_1$ and $\BA_5$ will be:
\[
\fontsize{8pt}{10pt}\selectfont
\begin{array}{rl}
\mathbf{A}_1 = &
\begin{array}{l}
\end{array} 
\left[
\begin{array}{*{4}{@{\hskip 0.05cm}c@{\hskip 0.05cm}}}\mathbf{a}_{12} & \mathbf{a}_{13} & \mathbf{a}_{14} & \mathbf{a}_{15} \\
1 & 0 & 0 & 0 \\
0 & 1 & 0 & 0 \\
0 & 0 & 1 & 0 \\
0 & 0 & 0 & 1 \\
\end{array}
\right] 
\end{array},
\fontsize{8pt}{10pt}\selectfont
\begin{array}{rl}
\mathbf{A}_5 = &
\begin{array}{l}
\end{array} 
\left[
\begin{array}{*{4}{@{\hskip 0.05cm}c@{\hskip 0.05cm}}}\mathbf{a}_{35} & \mathbf{a}_{25} & \mathbf{a}_{45} & \mathbf{a}_{15} \\
1 & 0 & 0 & 0 \\
0 & 1 & 1 & 0 \\
0 & 0 & 1 & 0 \\
1 & 0 & 0 & 1 \\
\end{array}
\right] 
\end{array}
\]

\begin{rem}
Each column of a user-specific coefficient matrix $\BA_k$ represents the sub-intervals during which the corresponding codeword is transmitted. Ideally, if each term is transmitted exactly once—resulting in an identity coefficient matrix—there is no noise amplification, as the noise is summed over the sub-intervals in which the desired codeword is transmitted. However, when the coefficient matrix deviates from this ideal structure, specific codewords must be repeated across multiple sub-intervals to ensure they are decodable from the system of equations \eqref{system_of_equations}. 
Such repetitions will penalize the user by increasing noise aggregation in the denominator of the SINR terms in \eqref{Non_convex_optimization}.
Since the algorithm constructs the coefficient matrix $\BA$ column-by-column in a greedy manner, user-specific $\BA_k$ matrices processed at a later stage will include more non-zero terms (more repetitions), and hence, they will be more penalized by the increased noise amplification.

To maximize fairness among users and mitigate the greedy nature of $\mathbf{A}$ construction, users are ordered in decreasing order based on their instantaneous channel conditions. The ordering algorithm starts with the strongest user, identified as the user with $\arg \max_k \|\mathbf{h}_k\|_2$. Subsequently, a successive projection method is applied to order the remaining users. Specifically, each remaining user is selected using $\arg \max_{k' \notin \hat{\mathcal{K}}} \mathbf{h}_{k'}^{H} \big(\mathbf{I} - \bar{\mathbf{H}}(\bar{\mathbf{H}}^{H} \bar{\mathbf{H}})^{-1} \bar{\mathbf{H}}^{H} \big) \mathbf{h}_{k'}$, where $\hat{\mathcal{K}}$ represents the set of already selected users, and $\bar{\mathbf{H}} = [\mathbf{h}_{\hat{\mathcal{K}}[1]}, \ldots, \mathbf{h}_{\hat{\mathcal{K}}[\lvert \hat{\mathcal{K}} \rvert]}]$. Users are then assigned to the coefficient matrix in decreasing priority order, ensuring that the most penalized user (by the coefficient matrix) has the largest channel gain, while the least penalized user has the highest spatial overlap with other users.
\end{rem}

%% file: Figs/Plot_4_javad.tex
    \resizebox{0.9\columnwidth}{!}{%
        \begin{tikzpicture}
            \begin{axis}
            [
                width = 1.2\columnwidth,
                height = 0.8\columnwidth,
                axis lines = center,
                xlabel near ticks,
                xlabel =  {\large SNR [dB]},
                ylabel =  {\large Symmetric Rate [bits/s]},
                ylabel near ticks,
                ymin = 0,
                ymax = 20,
                ytick = {0, 2, 4, 6, 8, 10, 12, 14, 16, 18, 20}, 
                xmin = 0,
                xmax = 50,
                xtick = {0, 5, 10, 15, 20, 25, 30, 35, 40, 45, 50},
                legend pos = north west,
                legend style = {font = \small},
                ticklabel style = {font = \large},
                grid=both,
                major grid style={line width=.2pt,draw=gray!30},
                grid style={line width=.1pt, draw=gray!10},
                minor tick num=4,
            ]

            \addplot
                 [red!80, line width=1.5pt]
                table[y=SIC,x=SNR]{Figs/data_Plot4_Javad.tex};
                \addlegendentry{\small SIC, optimized}

            \addplot
                [dashed, mark = star , red!80]
                table[y=SIC-ZF,x=SNR]{Figs/data_Plot4_Javad.tex};
                \addlegendentry{\small SIC, ZF}

            \addplot
                [blue!80, line width=1.5pt]
                table[y=Linear,x=SNR]{Figs/data_Plot4_Javad.tex};
                \addlegendentry{\small Proposed, optimized}

            \addplot
             [dashed, mark = star , blue!80]
             table[y=Linear-ZF,x=SNR]{Figs/data_Plot4_Javad.tex};
            \addlegendentry{\small Proposed, ZF}

            \addplot
                [green!80, line width=1.5pt]
                table[y=NO-CC,x=SNR]{Figs/data_Plot4_Javad.tex};
                \addlegendentry{\small No-CC, optimized}

            \addplot
                [dashed, mark = star , green!80]
                table[y=NO-CC-ZF,x=SNR]{Figs/data_Plot4_Javad.tex};
                \addlegendentry{\small No-CC, ZF}
            \end{axis}  
        \end{tikzpicture}
    }

%% file: Figs/Plot_3_javad.tex
     \resizebox{0.9\columnwidth}{!}{%
        \begin{tikzpicture}
            \begin{axis}
            [
                width = 1.2\columnwidth,
                height = 0.8\columnwidth,
                axis lines = center,
                xlabel near ticks,
                xlabel =  {\large SNR [dB]},
                ylabel =  {\large Symmetric Rate [bits/s]},
                ylabel near ticks,
                ymin = 0,
                ymax = 18,
                ytick = {0, 2, 4, 6, 8, 10, 12, 14, 16, 18}, 
                xmin = 0,
                xmax = 50,
                xtick = {0, 5, 10, 15, 20, 25, 30, 35, 40, 45, 50},
                legend pos = north west,
                legend style = {font = \large},
                ticklabel style={font=\large},
                grid=both,
                major grid style={line width=.2pt,draw=gray!30},
                grid style={line width=.1pt, draw=gray!10},
                minor tick num=4,
            ]

            \addplot
                [blue!80, line width=1.5pt]
                table[y=Linear,x=SNR]{Figs/data_Plot3_Javad.tex};
                \addlegendentry{\small Proposed, optimized}

            \addplot
             [dashed, mark = star , blue!80]
             table[y=Linear-ZF,x=SNR]{Figs/data_Plot3_Javad.tex};
            \addlegendentry{\small Proposed, ZF}

            \addplot
                 [red!80, line width=1.5pt]
                table[y=Signal-level,x=SNR]{Figs/data_Plot3_Javad.tex};
                \addlegendentry{\small Signal-level, ZF, power loading}

            \addplot
                [dashed, mark = star , red!80]
                table[y=Signal-level-eq,x=SNR]{Figs/data_Plot3_Javad.tex};
                \addlegendentry{\small Signal-level, ZF, equal power}


            \end{axis}

        \end{tikzpicture}
     }

%% file: Figs/Plot_7_javad.tex
    \resizebox{0.9\columnwidth}{!}{%
        \begin{tikzpicture}
            \begin{axis}
            [
                width = 1.2\columnwidth,
                height = 0.8\columnwidth,
                axis lines = center,
                xlabel near ticks,
                xlabel =  {\large SNR [dB]},
                ylabel =  {\large Symmetric Rate [bits/s]},
                ylabel near ticks,
                ymin = 0,
                ymax = 4,
                ytick = {0, 1, 2, 3, 4}, 
                xmin = 0,
                xmax = 40,
                xtick = {0, 5, 10, 15, 20, 25, 30, 35, 40},
                legend pos = north west,
                legend style = {font = \large},
                ticklabel style={font=\large},
                grid=both,
                major grid style={line width=.2pt,draw=gray!30},
                grid style={line width=.1pt, draw=gray!10},
                minor tick num=4,
            ]


            \addplot
                [blue!80, line width=1.5pt]
                table[y=Linear,x=SNR]{Figs/data_Plot7_Javad.tex};
                \addlegendentry{\small Propozed, optimized}

            \addplot
                [red!80, line width=1.5pt]
                table[y=ZF-pl,x=SNR]{Figs/data_Plot7_Javad.tex};
                \addlegendentry{\small Signal-level, ZF, power loading}

            \end{axis}

        \end{tikzpicture}
    }

%% file: Figs/Plot_5_javad.tex
     \resizebox{0.9\columnwidth}{!}{%
        \begin{tikzpicture}[scale=0.66]
            \begin{axis}
            [
                width = 1.2\columnwidth,
                height = 0.8\columnwidth,
                axis lines = center,
                xlabel near ticks,
                xlabel =  {\large SNR [dB]},
                ylabel =  {\large Symmetric Rate [bits/s]},
                ylabel near ticks,
                ymin = 0,
                ymax = 16,
                ytick = {0, 2, 4, 6, 8, 10, 12, 14, 16}, 
                xmin = 0,
                xmax = 50,
                xtick = {0, 5, 10, 15, 20, 25, 30, 35, 40, 45, 50},
                legend pos = north west,
                legend style = {font = \large},
                ticklabel style={font=\large},
                grid=both,
                major grid style={line width=.2pt,draw=gray!30},
                grid style={line width=.1pt, draw=gray!10},
                minor tick num=4,
            ]

             \addplot
                 [black!80, line width=1.5pt]
                table[y=Greedy,x=SNR]{Figs/data_Plot5_Javad.tex};
                \addlegendentry{\small Sparse, Successive Projection}

            \addplot
                [blue!80, line width=1.5pt]
                table[y=Linear,x=SNR]{Figs/data_Plot5_Javad.tex};
                \addlegendentry{\small Sparse}

            \addplot
             [brown!80, line width=1.5pt]
             table[y=Linear-eq,x=SNR]{Figs/data_Plot5_Javad.tex};
            \addlegendentry{\small Equal-Distance}

            \addplot
                 [red!80, line width=1.5pt]
                table[y=Linear-rand,x=SNR]{Figs/data_Plot5_Javad.tex};
                \addlegendentry{\small Random}
            
            \end{axis}

        \end{tikzpicture}
     }

%% file: SimulationResults.tex
\section{Simulation Results}
\label{section:Simulations}
We use numerical simulations to compare our proposed linear solution with other schemes. The channels are considered i.i.d complex Gaussian, and the SNR is defined as $\frac{P_T}{N_0}$, where $P_T$ is the power budget at the
BS and $N_0$ denotes the fixed noise variance.
%
In Fig.~\ref{fig:SIC_5_user}, we compare how our proposed scheme coupled with the sparse matrix design performs against the baseline non-linear solution in~\cite{tolli2017multi} (shown as SIC in the figure) and another baseline, dubbed as No-CC, where the cache memories are used only for the local caching gain and no CC technique is applied (users are selected using a cyclic shift operation and are served by spatial multiplexing). For all three schemes, both optimized and ZF beamformers are compared. As can be seen, our linear solution causes a loss compared to SIC, as the SIC-free decoding imposes more restrictions on the transmission design. However, it outperforms No-CC by a large margin (the gap will further increase for larger $t$).

Fig.~\ref{fig:Signal_level} compares our proposed scheme coupled with the sparse matrix design with another baseline: the signal-level CC scheme with ZF beamforming~\cite{shariatpanahi2017multi}. As can be seen, even with ZF beamforming, our SIC-free solution outperforms the signal-level solution. This is because of the superior multicasting gain in our scheme, as the missing subpackets are transmitted in XOR codewords of size $t+1$. For a better comparison, we have included two other cases in the figure: our proposed solution with optimized beamformers, which outperforms all the other curves, and the signal-level scheme with ZF with equal power allocation (no power loading), which has the worst performance among all.

Fig.~\ref{fig:20-users} demonstrates the scalability of the proposed scheme as the number of users increases. Note that the SIC strategy is excluded from the plot as it becomes impractical in large networks due to the exponentially increasing number of SINR and sum-rate constraints and the complexity of beamforming design. In contrast, the scheduling mechanism of the proposed sparse strategy simplifies the beamforming design, while our linear beamforming optimization framework in~\eqref{Non_convex_optimization} substantially reduces the number of SINR and sum-rate constraints. The notable gap between the proposed solution and the signal-level scheme highlights the beamforming gain achieved by the proposed bit-level scheme and the effectiveness of the user scheduling. This gap shows the inferior performance of signal-level schemes in the finite-SNR regimes~\cite{salehi2022multi,salehi2022enhancing}.

Finally, in Fig.~\ref{fig:coefficient_matrix}, we investigate the effect of the coefficient matrix design. Optimized beamformers are employed in all cases. As can be seen, the sparse design outperforms both equal-distance and random designs, and the random design has the worst output. This is expected as the sparse design has the best structure that tries to maximize the signal and minimize the interference power simultaneously in~\eqref{Non_convex_optimization}. The sparse design, combined with the successive projection-based user ordering, outperforms all the coefficient matrix designs. 

%% file: conclusions.tex
\section{Conclusion}
\label{sec:conclusions}
{We proposed a novel linear transmission solution based on the repetition of the signal terms with varying linearly independent sets of coefficients to eliminate the need for the complex SIC structure while keeping the underlying bit-domain multicasting gain of coded caching. 
Different strategies were considered to build the coefficient matrix, and the effect of the coefficient matrix on performance was investigated.
Numerical simulations were used to compare the proposed scheme with the state of the art.
}